\documentclass[conference]{IEEEtran}
\IEEEoverridecommandlockouts
\usepackage{cite}
\usepackage{amsmath,amssymb,amsfonts}
\usepackage{algorithmic}
\usepackage{graphicx}
\usepackage{textcomp}
\usepackage{xcolor}
\def\BibTeX{{\rm B\kern-.05em{\sc i\kern-.025em b}\kern-.08em
    T\kern-.1667em\lower.7ex\hbox{E}\kern-.125emX}}

\usepackage{caption}
\usepackage{url}
\usepackage{hyperref}
\hypersetup{hidelinks}
\usepackage{xurl}

\begin{document}

% \title{Conference Paper Title*\\
% {\footnotesize \textsuperscript{*}Note: Sub-titles are not captured for https://ieeexplore.ieee.org  and
% should not be used}
% \thanks{Identify applicable funding agency here. If none, delete this.}
% }
% 
% \author{\IEEEauthorblockN{1\textsuperscript{st} Given Name Surname}
% \IEEEauthorblockA{\textit{dept. name of organization (of Aff.)} \\
% \textit{name of organization (of Aff.)}\\
% City, Country \\
% email address or ORCID}
% \and
% \IEEEauthorblockN{2\textsuperscript{nd} Given Name Surname}
% \IEEEauthorblockA{\textit{dept. name of organization (of Aff.)} \\
% \textit{name of organization (of Aff.)}\\
% City, Country \\
% email address or ORCID}
% \and
% \IEEEauthorblockN{3\textsuperscript{rd} Given Name Surname}
% \IEEEauthorblockA{\textit{dept. name of organization (of Aff.)} \\
% \textit{name of organization (of Aff.)}\\
% City, Country \\
% email address or ORCID}
% \and
% \IEEEauthorblockN{4\textsuperscript{th} Given Name Surname}
% \IEEEauthorblockA{\textit{dept. name of organization (of Aff.)} \\
% \textit{name of organization (of Aff.)}\\
% City, Country \\
% email address or ORCID}
% \and
% \IEEEauthorblockN{5\textsuperscript{th} Given Name Surname}
% \IEEEauthorblockA{\textit{dept. name of organization (of Aff.)} \\
% \textit{name of organization (of Aff.)}\\
% City, Country \\
% email address or ORCID}
% \and
% \IEEEauthorblockN{6\textsuperscript{th} Given Name Surname}
% \IEEEauthorblockA{\textit{dept. name of organization (of Aff.)} \\
% \textit{name of organization (of Aff.)}\\
% City, Country \\
% email address or ORCID}
% }

\title{Spatial-Temporal Hybrid Path Planning for Multi-UAV Methane Emission Rate Quantification}

\author{\IEEEauthorblockN{Ruikun Wu\IEEEauthorrefmark{1}, Derek Hollenbeck\IEEEauthorrefmark{2} and Sihua Shao\IEEEauthorrefmark{1}} \IEEEauthorblockA{\IEEEauthorrefmark{1}Department of Electrical Engineering,
Colorado School of Mines, Golden, Colorado 80401, USA\\Emails: ruikun\_wu@mines.edu, sihua.shao@mines.edu
\\\IEEEauthorrefmark{2}Department of Mechanical Engineering, University of California Merced, Merced, California 95343, USA\\Email: dhollenbeck@ucmerced.edu}
\vspace{-20pt}
\thanks{The work was supported by the Colorado Office of Economic Development \& International Trade (OEDIT) under Award no. CTGG1 2026-3294. (\textit{Corresponding author: Sihua Shao.})}
}

\maketitle

\begin{abstract}
Accurate emission-rate quantification of methane point sources is essential for inventory verification and leak detection. UAV-borne sensing provides high-resolution, near-source measurements, but existing quantification methods mostly follow a predefined trajectory and rely on a single UAV. Building on a source receptor matrix (SRM) estimator and empirical observability Gramian (EMGR)-based path planning, this paper proposes a spatial-temporal hybrid path planning strategy for multi-UAV methane emission rate quantification. Measurements are time-averaged, and the forward Gaussian plume model is calibrated against a filament-based plume
simulator (MOABS) to reduce model mismatch. Cooperating UAVs first spread out spatially to locate the source quickly, then switch to sequentially retracing a leader trajectory with a time lag to reduce quantification variance. Simulations show that the spatial stage localizes the source markedly faster than a single-UAV
baseline, the temporal stage yields the lowest variance but converges slowly, and the proposed hybrid strategy converges substantially faster than the purely temporal strategy while matching its steady-state accuracy, with a narrower error band than the purely spatial strategy. A window-length study further shows that longer time-averaging windows exhibit diminishing returns, justifying the balance between spatial coverage and temporal averaging.
\end{abstract}


\section{Introduction}
% Should follow these five aspects:
% Background 
% Intention 
% Method
% Results
% Conclusion
Methane is a potent greenhouse gas, and oil and gas supply chains
are among its largest anthropogenic sources. Facility-scale
measurements indicate that U.S. supply-chain emissions are about
$60\%$ higher than the inventory estimate, because inventory
methods miss emissions released under abnormal operating
conditions~\cite{Alvarez2018}. Satellite observations show that a
small number of ultra-emitters alone account for $8-12\%$ of global
oil and gas production emissions~\cite{Lauvaux2021}. Producers and
regulators therefore need to quantify the emission rate of
individual sources reliably, quickly and at low cost.

Several measurement platforms serve this purpose, each trading
coverage for resolution: satellite point-source imagers cover large
areas but have detection thresholds of roughly
$100$--$10{,}000$\,kg/h~\cite{Jacob2022}; airborne LiDAR gas mapping
(e.g., Bridger Photonics) operates at facility-to-basin scales but
degrades as wind speed increases~\cite{Johnson2021}; and optical
gas imaging (OGI) cameras resolve individual leaks with a median
detection limit of about $20$\,g/h at $6$\,m, but require an
on-site operator~\cite{Ravikumar2018}. UAVs carrying in-situ or
path-integrated sensors fill the gap: they deploy rapidly, sample
close to the source at high resolution, and can be
automated~\cite{Burgues2020,Ravikumar2019}. The remaining
difficulty is algorithmic: deciding where to fly and converting
measurements into an accurate emission rate.

Two families of methods convert UAV measurements into emission
rates: mass balance (MB), which integrates measured flux through a
downwind sampling surface~\cite{Conley2017,Yang2018,Golston2018},
and near-field Gaussian plume inversion (NGI), which fits a plume
model to the measurements~\cite{Shah2019,Shah2020}. Both are
typically flown on predefined patterns (vertical planes, lawnmower
paths) that cannot adapt to plume changes such as a sudden wind
shift, and their accuracy depends on how well the fixed pattern
covers the plume: short flights introduce a systematic flux
bias~\cite{Shah2019}, and manual plume-following, tried as a
remedy, introduces its own bias~\cite{Shah2020}.

Informative path planning offers a principled alternative, in which
the UAV selects the next waypoint by maximizing the expected
information gain about the quantity of interest~\cite{Popovic2020}.
For airborne releases, information-based search strategies estimate
the source term online and steer a mobile sensor toward informative
measurements~\cite{Hutchinson2018SourceTE}. Most relevant to this
work, Hollenbeck et al.~\cite{Hollenbeck2025AnOM} combine an SRM
estimator, solved recursively, with a Gaussian plume digital twin
and use the EMGR to choose sensing locations in real time. This
approach has three limitations that we address here: (i) it was
evaluated only in simulations where the environment was itself the
Gaussian plume model, so no model mismatch was present, and a
fixed-path field test revealed a substantial mismatch-driven bias;
(ii) the forward model is static and time-invariant, whereas real
plumes are turbulent and intermittent; and (iii) a single UAV must
both locate the source and dwell long enough for valid measurements,
two conflicting objectives.

This paper extends the SRM/EMGR framework
of~\cite{Hollenbeck2025AnOM} to cooperating multiple UAVs in a
time-varying plume. The contributions are as follows:
\begin{itemize}
\item We integrate the quantification pipeline with MOABS, a
filament-based simulator producing intermittent, time-varying
plumes, and calibrate the Gaussian plume forward model against its
time-averaged field to reduce model mismatch.

\item We propose a spatial-temporal hybrid path planning strategy
for multiple UAVs, combining spatial diversity for fast source
localization with temporal diversity for low-variance
quantification.

\item We study how the time-averaging window length affects
quantification accuracy, showing that longer windows reduce
variance but yield rapidly diminishing improvements in mean error.
\end{itemize}

Our implementation is publicly available on Github.\footnote{\url{https://github.com/Shao-wireless-lab/Multi-drone-source-seeking-and-quantification-.git}}

% \textit{Organization:} Sec \ref{sec_sysmodel} presents the problem statement and the quantification pipeline. Sec \ref{sec_proposed_path_planning} describes the time-averaged measurement model and the proposed hybrid path planning strategy. Sec \ref{sec:PERFORMANCE_EVALUATION} reports the simulation results, and Sec \ref{sec:Ch5-Conclusion} concludes the paper with future work.

%%%%%%%%%%%%%%%%%%%%%%%%%%%%%%%%%%%%%%%%%%%%%%%%%%%%%%%%%%
\section{System Model}
\label{sec_sysmodel}
\subsection{Problem Statement} \label{sec_prob_state}

Methane emission-rate estimation is formulated as a partially
observed inverse problem: a single point-source emission on the
ground has unknown rate $Q$ (kg/s) and unknown location
$x_s\in \mathbb{R}^2$ within a bounded area $\Omega_2 \subset
\mathbb{R}^2$. UAVs are restricted to a fixed altitude since our
source receptor matrix estimator (Sec.~\ref{sec_SRM}) operates in
2D. A UAV equipped with methane and wind sensors traverses
$\Omega_2$, collecting measurements $\{y_k\}_{k=1}^K$ at positions
$\{x_m(k)\}_{k=1}^K$ with concurrent wind estimates, and the goal is
to recover $\hat{Q}$ that best explains them under a fast-running
forward model. This accuracy depends not only on the
source-receptor system but on \textit{where} and \textit{when} the
UAV samples the plume, motivating the quantification pipeline
(Secs.~\ref{sec_GPM}--\ref{sec_EMGR}, based
on~\cite{Hollenbeck2025AnOM}) and path planning strategy
(Sec.~\ref{sec_proposed_path_planning}) developed below.

\subsection{Gaussian Plume Forward Model} \label{sec_GPM}
Real-time inversion requires a fast forward model, so we adopt the
steady-state Gaussian plume model (GPM). Given a measurement
location $\mathbf{x}=[x_1,x_2,x_3]$ in the plume-local frame
(rotated to the wind direction and translated to the source, with
$x,y,z$ denoting the downwind, crosswind, and vertical directions), the predicted concentration is expressed as
\begin{equation}
y_m = \mathcal{M}(\mathbf{x}, \boldsymbol{\theta}) = \frac{Q}{U} D_2(\mathbf{x}, \boldsymbol{\theta})\, D_3(\mathbf{x}, \boldsymbol{\theta}) \label{eq1}
\end{equation}
\begin{equation}
D_2(\mathbf{x}, \boldsymbol{\theta}) = \frac{1}{\sqrt{2\pi}\sigma_2}
    \exp\!\left(\frac{-(x_2-\mu_2)^2}{2\sigma_2^2}\right) \label{eq_y_dispersion}
\end{equation}
% \begin{equation}
%     D_3(\mathbf{x}, \theta) = \frac{1}{\sqrt{2\pi}\sigma_3}
%         \left[\exp\!\left(\frac{-(x_3-H)^2}{2\sigma_3^2}\right)
%         + \exp\!\left(\frac{-(x_3+H)^2}{2\sigma_3^2}\right)\right] \label{eq_z_dispersion}
% \end{equation}
\begin{align}
D_3(\mathbf{x}, \boldsymbol{\theta}) = \frac{1}{\sqrt{2\pi}\sigma_3}&
    \bigg[\exp\!\left(\frac{-(x_3-H)^2}{2\sigma_3^2}\right) \notag\\
    &+ \exp\!\left(\frac{-(x_3+H)^2}{2\sigma_3^2}\right)\bigg]
    \label{eq_z_dispersion}
\end{align}
where $\boldsymbol{\theta}=[Q,U,\mu_2,\sigma_2,\sigma_3, \cdot]$, $\textit{U}$ is the mean wind speed, $\textit{H}$ is the source release height (here $\textit{H}=0$ for a ground-level source), $\mu_2$ is the coordinate along $y$ direction,  $\sigma_2$ and $\sigma_3$ are plume dispersion parameters along $y$ and $z$ direction respectively. One approach of determining the dispersion parameters is using Pasquill-Gifford atmospheric stability class~\cite{Davidson1990}, which characterizes the dispersion parameters as power laws of the downwind distance $x_1$,
\begin{equation}
    \sigma_2(x_1) = a_2 x_1^{b_2}, \qquad
    \sigma_3(x_1) = a_3 x_1^{b_3},
    \label{eq:sigma_powerlaw}
\end{equation}
Rather than adopting literature values for a fixed stability class, the coefficients $(a_2,b_2,a_3,b_3)$ in this work are calibrated using time-averaged data generated by filament-based dispersion simulator, which is called Methane Odor Abatement Simulator (MOABS). After calibration, the forward model shape can match the time-averaged turbulence of our simulated environment.

\subsection{Source Receptor Matrix and Recursive Estimation} \label{sec_SRM}
To enable real-time inversion, we need an efficient approach to solve the inverse problem. We adopt the source-receptor matrix (SRM)~\cite{Seibert2004SourcereceptorMC}
to solve the inverse problem efficiently online; for a Gaussian
plume the SR relationship is linear. The receptor is the UAV's
sensor reading, and candidate sources are placed at $N_q$
discretized cells $X_q=[\textbf{x}_q^{(1)T},\cdots,
\textbf{x}_q^{(N_q)T}]^T \subseteq \Omega_2$ spanning the domain.
Given $k$ measurement locations
$\mathbb{X}_m=[\textbf{x}_m(1),\cdots,\textbf{x}_m(k)]^T$ and
assuming unit source rate at each candidate, GPM evaluated at each
measurement location builds an SR matrix $H\in\mathbb{R}^{k\times
N_q}$, with entries
\begin{equation}
H_{ij} = \frac{1}{U(i)}\,
    \mathcal{M}\big(\mathbf{x}_m(i), \boldsymbol{\theta}_q^{(ij)}\big)
\end{equation}
where $\theta_q^{(ij)}$ is evaluated in the local frame relative to
candidate source $j$, using the calibrated $\sigma_2, \sigma_3$
from equation \eqref{eq:sigma_powerlaw}. The predicted measurement
vector is then $\hat{\mathbf{Y}} = \mathbf{H}\hat{\mathbf{Q}}$, and
the regularized least-squares estimate is
\begin{equation}
    \hat{\mathbf{Q}} = \arg\min_{\mathbf{Q}}
        \|\mathbf{Y} - \mathbf{H}\mathbf{Q}\|_2^2
        + \beta \|\mathbf{Q}\|_2^2 . \label{eq_Q_hat}
\end{equation}
For online operation, \eqref{eq_Q_hat} is solved recursively via the
standard recursive least-squares (RLS) update, initialized with
$P_0 = \beta^{-1}I$, $\hat{\mathbf{Q}}_0 = \mathbf{0}$.

Once the estimate stabilizes near an anchor point $\mathbf{x}_a$
(Sec.~\ref{sec_anchor}), the final readout is the maximum of
$\hat{\mathbf{Q}}_k$ within a neighborhood of $\mathbf{x}_a$ rather
than the raw argmax over the full domain, mitigating the
mass-spreading bias from the $\ell_2$ regularization
in~\eqref{eq_Q_hat} (unlike~\cite{Hollenbeck2025AnOM}, where the
anchor instead redirects the UAV when the observability-based
controller fails to converge).

\subsection{Empirical Observability Gramian}    \label{sec_EMGR}
% 排放源量化问题定义：Q_true反演，域大小200×200m，源位置未知/给定
% 甲烷浓度观测模型（Gaussian plume forward model，简述sigma_y/sigma_z幂律形式）
% SRM + RLS估计框架简述
To quantify how informative a given (or candidate) UAV trajectory
is for the estimation problem in Section \ref{sec_SRM}, we adopt the
empirical observability Gramian (EMGR)~\cite{Lall1999}. For a
nonlinear system $\dot{x} = f(t,x(t),u(t),\theta)$, $y(t) =
h(t,x(t),u(t),\theta)$, the EMGR is
\begin{equation}
    \hat{W}_O = \frac{1}{|S_x|}\sum_{l=1}^{|S_x|} \frac{1}{d_l^2}
        \int_0^\infty \Psi^l(t)\, dt,
\end{equation}
where $\Psi^l = (y^{l_i}(t) - \bar{y}^{l_i})^T(y^{l_i}(t) -
\bar{y}^{l_i})$ is computed from perturbed initial conditions
$x_0^{l_i} = d_l \epsilon^i + \bar{x}$. Following~\cite{Hollenbeck2025AnOM},
we combine the unobservability index and D-optimality of
$\hat{W}_O$ into a single scalar information metric,
\begin{equation}
    J(\mathbf{x}) = \alpha\, \lambda_{\min}^{-1}(\hat{W}_O)
        - \log\det(\hat{W}_O),
    \label{eq:emgr_cost}
\end{equation}
which is minimized at candidate sensing locations to guide
information-seeking motion. We denote this quantity evaluated at a
candidate waypoint $\mathbf{p}$ as $I_{\text{EMGR}}(\mathbf{p})
\triangleq J(\mathbf{p})$, i.e., lower values indicate lower uncertainty and thus more informative locations; this term is reused directly as one component of the weighted score in the proposed path planning strategy.

Together, Sections \ref{sec_GPM} to \ref{sec_EMGR} define a complete pipeline from a
measurement trajectory to an emission-rate estimate: the forward
model relates candidate sources to predicted concentrations, the
SRM/RLS framework inverts this relation online, and the EMGR
provides a principled measure of how much a given trajectory
segment contributes to reducing estimation uncertainty. However,
this pipeline treats the measurement trajectory as given; it does
not by itself specify how the UAV should move to jointly
balance spatial coverage, temporal dwell time, and information
gain under a limited flight budget. This gap motivates the
spatial--temporal hybrid path planning strategy introduced next.

%%%%%%%%%%%%%%%%%%%%%%%%%%
\section{Proposed Path Planning Strategies} \label{sec_proposed_path_planning}

\subsection{Time-averaged Measurement and GP Joint Optimization} \label{sec_time_avg_gp}
Instantaneous methane concentration measurements are sparse and
intermittent, making them difficult to use directly for
quantification; we therefore time-average them first. A UAV that
dwells at a sampling location for a non-negligible period $\tau$
produces a time-averaged concentration reading,
\begin{equation}
    y_k = \frac{1}{\tau}\int_{t_k}^{t_k+\tau}c(\textbf{x}_m(k),t)dt \label{eq_time_avg}
\end{equation}
which also suppresses turbulence-driven fluctuations, making the
RLS inversion of Sec.~\ref{sec_SRM} better conditioned.

Since the forward GP model is static while MOABS is dynamic and
time-variant, a model mismatch exists that degrades quantification
accuracy. To reduce it, we calibrate the GP model against a static
target obtained by globally time-averaging the full MOABS dataset.
Since the forward-model shape is directly determined by the
dispersion parameters $(a_2,b_2,a_3,b_3)$ in~\eqref{eq:sigma_powerlaw},
we fit these against the global time-averaged field rather than
adopting values for a fixed stability class. We additionally extend the $y$-direction term in~\eqref{eq_y_dispersion} to admit a downwind-dependent offset,
\begin{equation}
    \begin{cases}
            x_{\text{eff}} = x_1 + x_0 \\
            y_{\text{center}}(x_{\text{eff}}) = y_{c0} + y_{c,\text{slope}}x_{\text{eff}} 
    \end{cases}   \label{eq_y_center}
\end{equation}

where $x_0$ is GP source location offset along $x$ direction, $x_{\text{eff}}$ is the calibrated distance from the sampling point to the GP source along $x$ direction. Similarly, $y_{c0}$ is the $y$-direction offset and $y_{c,\text{slope}}$ is slope coefficient of $x_{\text{eff}}$. Therefore, $D_2$ is evaluated at $(x_2 - y_{\text{center}})$ rather
than $x_2$ directly. The additional degree of freedom in  \eqref{eq_y_center} allows the calibrated shape to absorb any systematic crosswind drift present in a given simulated or real environment, rather than forcing the plume centerline to remain fixed at the source's $y$ coordinate. Section~\ref{sec:PERFORMANCE_EVALUATION} shows these terms are in fact necessary for our simulation environment.

All shape parameters, including $y_{c0}$ and $y_{c,\text{slope}}$, are obtained by a single joint optimization before deployment:
\begin{equation}
\boldsymbol{\theta}^* = \mathop{argmax}_{\boldsymbol{\theta}} R^2(\bar c_{emp}, \mathcal{M}(\cdot,\boldsymbol{\theta}))
\end{equation}
where $\bar c_{emp}$ is the global time-averaged concentration field over the full duration of MOABS dataset, and $R^2$ measures the goodness of fit between this global time-averaged concentration field and the field of the GP model. Calibrating these shape parameters from this long-horizon fit decouples two design questions: what shape the forward model should take, and how many seconds a UAV should dwell at each sampling location during online quantification.

The latter question is essentially a trade-off: an overly short time-average window $\tau$ yields noisy and poorly conditioned measurements, while an overly long window reduces the number of valid independent measurements and makes the quantification algorithm inefficient. We treat $\tau$ as a design parameter to be chosen empirically in Section \ref{sec:PERFORMANCE_EVALUATION}.

\subsection{Spatial-temporal Hybrid Multi-UAV Path Planning} \label{sec_hybrid_planning}
A single UAV following an information-driven path
(Sec.~\ref{sec_EMGR}) must pursue two competing objectives:
covering enough of the domain to localize the source, and dwelling
long enough at each location to obtain a valid time-averaged
measurement. We resolve this tension by decomposing the problem
across $N$ cooperating UAVs into two complementary forms of
diversity: \textit{spatial diversity}, where UAVs explore different regions in parallel to accelerate localization, and \textit{temporal diversity}, where agents revisit the same locations at different times to accumulate information on the turbulent plume, reducing quantification variance through the RLS update's ensemble-averaging effect.

\subsubsection{Anchor Point}  \label{sec_anchor}
Both stages are coordinated through a shared anchor point
$\textbf{x}_a$, an online smoothed estimate of the source location,
following Algorithm 1 of our previous
work~\cite{Li2026MultiAgentRL}: when a measured concentration at
$\textbf{x}_i$ exceeds threshold $\rho_{\text{th}}$ and the vector
$\Delta \textbf{x}(t)=\textbf{x}_i - \textbf{x}_a$ falls within the
field-of-view boundary around the upwind direction $-\textbf{V}_\text{wind}$,
the anchor is updated to $\textbf{x}_i$. Whereas prior work used the
anchor to trace the plume, here it serves two roles: (i)
initializing and informing the path-planning score below, and (ii)
determining the readout neighborhood for the final quantification
estimate in Sec.~\ref{sec_SRM}.

\subsubsection{Stage 1 -- Spatial Diversity Planning}
Prior to anchor point convergence, all $N$ agents independently select their next waypoint $\textbf{x}_p$ from a discrete action set by minimizing a weighted score
\begin{align}
    S_{\text{spatial}}(\mathbf{x}_p) &= w_1 I_{\text{EMGR}}(\mathbf{x}_p)
        + w_2 D_{\text{anchor}}(\mathbf{x}_p)
        + w_3 P_{\text{bound}}(\mathbf{x}_p) \notag \\
        & \ \ \ + w_4 P_{\text{visit}}(\mathbf{x}_p)
        + w_5 P_{\text{sep}}(\mathbf{x}_p)  \label{eq_score_spatial}
\end{align}
where $I_{\text {EMGR}}$ is the EMGR-based information term of section \ref{sec_EMGR} and $w_i$ is the weight of each score term respectively. The remaining four terms are defined as follows. 

\emph{Anchor attraction.} Once an anchor $\textbf{x}_a$ has been established, each agent is pulled toward it via 
\begin{equation}
    D_{\text{anchor}}(\textbf{x}_p) = \|\textbf{x}_p-\textbf{x}_a\|_2 \label{eq_anchor_term}
\end{equation}
so that minimizing \eqref{eq_score_spatial} favors waypoints closer to the current estimate of source location. This term is omitted before anchor is established.

\emph{Boundary penalty.} To discourage agents from hitting the edges of the bounded domain $\Omega_2 \subset[0,L]^2$, we develop a penalty term for each of the four boundaries,
%\begin{equation}
%        P_{\text{bound}}(\mathbf{x}_p) = \frac{1}{x_{p,1}+1}
%        + \frac{1}{L - x_{p,1}+1}
%        + \frac{1}{x_{p,2}+1}
%        + \frac{1}{L - x_{p,2}+1},
%    \label{eq_boundary_term}
%\end{equation}
\begin{multline}
    P_{\text{bound}}(\mathbf{x}_p) = \frac{1}{x_{p,1}+1}
        + \frac{1}{L - x_{p,1}+1} \\
        + \frac{1}{x_{p,2}+1}
        + \frac{1}{L - x_{p,2}+1},
    \label{eq_boundary_term}
\end{multline}
where $x_{p,1}, x_{p,2}$ are the coordinates of $\textbf{x}_p$;
each term grows large near the corresponding edge and stays small
in the interior.

\emph{Revisit penalty.} To promote exploration and discourage agents from repeatedly visiting previously sampled grids, we use the same 2D grids with MOABS and penalize candidates by the accumulated visit count of their corresponding grid,
\begin{equation}
        P_{\text{visit}}(\mathbf{x}_p) = V\big(\text{grid}(\mathbf{x}_p)\big),
    \label{eq_visit_term}
\end{equation}
where grid($\cdot$) maps a continuous position to its discretized grid index and $V$ is incremented each time an agent visits that grid.

\emph{Separation penalty.} To encourage spatial diversity among the $N$ agents, each agent penalizes candidate waypoints that would collide with other agents. For a specific agent, the separation score between its next waypoint $\textbf{x}_p$ and other $N-1$ agents' current positions is calculated as follows:
\begin{equation}
    P_{\text{sep}}(\mathbf{x}_p) = \sum_{j}
        \max\big(0,\; d_{\text{safe}} - \|\mathbf{x}_p - \mathbf{x}_j\|_2\big),
    \label{eq_sep_term}
\end{equation}
where $\mathbf{x}_j$ ranges over the other $N-1$ agents' positions; the term is zero beyond $d_{\text{safe}}$ and grows linearly as
agents approach or overlap. Together
with~\eqref{eq_anchor_term}--\eqref{eq_visit_term}, this lets the
$N$ agents collectively explore the domain and converge on the
anchor faster than a single agent could.

\subsubsection{Stage 2 -- Temporal Diversity Planning}
If the movement of location of anchor is smaller than predefined
threshold $d_\text{stable}$ for $k_s$ consecutive times, the anchor is considered to be converged. Once the anchor has converged, the planning strategy switches to a temporal-diversity mode while the intention also switches from source localization to emission quantification. In this mode, a single reference (``leader'') trajectory is generated using the same weighted score
as~\eqref{eq_score_spatial}, but with the separation penalty
$P_{\text{sep}}$ dropped, since revisiting the same location is now desirable rather than penalized.

The remaining $N-1$ agents do not plan independently; each simply
follows the same leader trajectory with a delay of
$i\Delta t_\text{lag}$ steps for agent $i$. Although agent $i$'s
position at step $k$ is copied from the leader's history, its
measurement is a fresh sample taken at the current time $k$, so the same physical location is effectively sampled $N$ times over the mission. This extends the effective time-averaging window, reducing model mismatch between the calibrated forward GP and MOABS and improving quantification accuracy. Therefore, temporal diversity is achieved by sequential following planning.

The lag $\Delta t_\text{lag}$ governs a trade-off between quantification stability and efficiency. Larger $\Delta t_\text{lag}$ allows longer time average window and more quantification stability, while smaller $\Delta t_\text{lag}$ keeps the quantification more efficient. We select this parameter empirically in Section \ref{sec:PERFORMANCE_EVALUATION}.

\subsubsection{Hybrid Strategy}
The proposed  hybrid strategy composes the two stages above: $N$ agents execute the spatial warmup of Stage 1 until the anchor stabilizes, then switch to the temporal sequential-sensing of Stage 2 for the rest of the mission, combining efficient localization with variance-reducing dwell time. This is evaluated in Section~\ref{sec:PERFORMANCE_EVALUATION} against single-agent,
purely spatial, and purely temporal baselines.

%%%%%%%%%%%%%%%%%%%%%%%%%%%%%%%%%%%%%%%%%%%%%%%%%%%%%%%%%%%%%%%%%%%%%%%
\section{Performance evaluation}
\label{sec:PERFORMANCE_EVALUATION}
\subsection{Experiment Setup}
\label{sec:exp_setup}
Simulation data are generated by Methane Odor Abatement Simulator (MOABS), a filament-based dispersion simulator described in detail in our previous work~\cite{Li2026MultiAgentRL}. We have validated that the statistical pattern of MOABS-generated concentration fields closely resembles real-world flight data
provided by Project Canary, a third-party methane monitoring
provider.

MOABS propagates a population of Lagrangian filaments from a
fixed ground-level point source through a turbulent flow field,
producing concentration and wind fields with intermittent,
non-Gaussian plume structure. A source of true rate
$Q_{\text{true}}=5.225\times10^{-5}$ kg/s is placed at $(80,60)$\,m within a $200\,\text{m}\times200\,\text{m}$ domain $\Omega_2$, with concentration and wind recorded on a $100\times100$ grid at
$5$\,m height over a $10{,}000\,\text{s}$ run. This dataset serves as both the calibration source (Section~\ref{sec_time_avg_gp}) and the evaluation environment for all strategies.

Unless noted otherwise, all path-planning experiments use $N=3$ UAVs and a candidate-source grid of $25\times 25$ cells over $\Omega_2$. Strategy-specific parameters (weights $w_i$, thresholds $\rho_{\text{th}}, d_\text{stable}, d_\text{safe}$, and the sequential-following lag $\Delta t_\text{lag}$) are held fixed except for parameter sweeping, as in Section \ref{sec:window_length_result}.

\begin{figure}[!t]
    \centering
    \begin{minipage}[b]{3.5in}
        \centering
        \includegraphics[width=\linewidth]{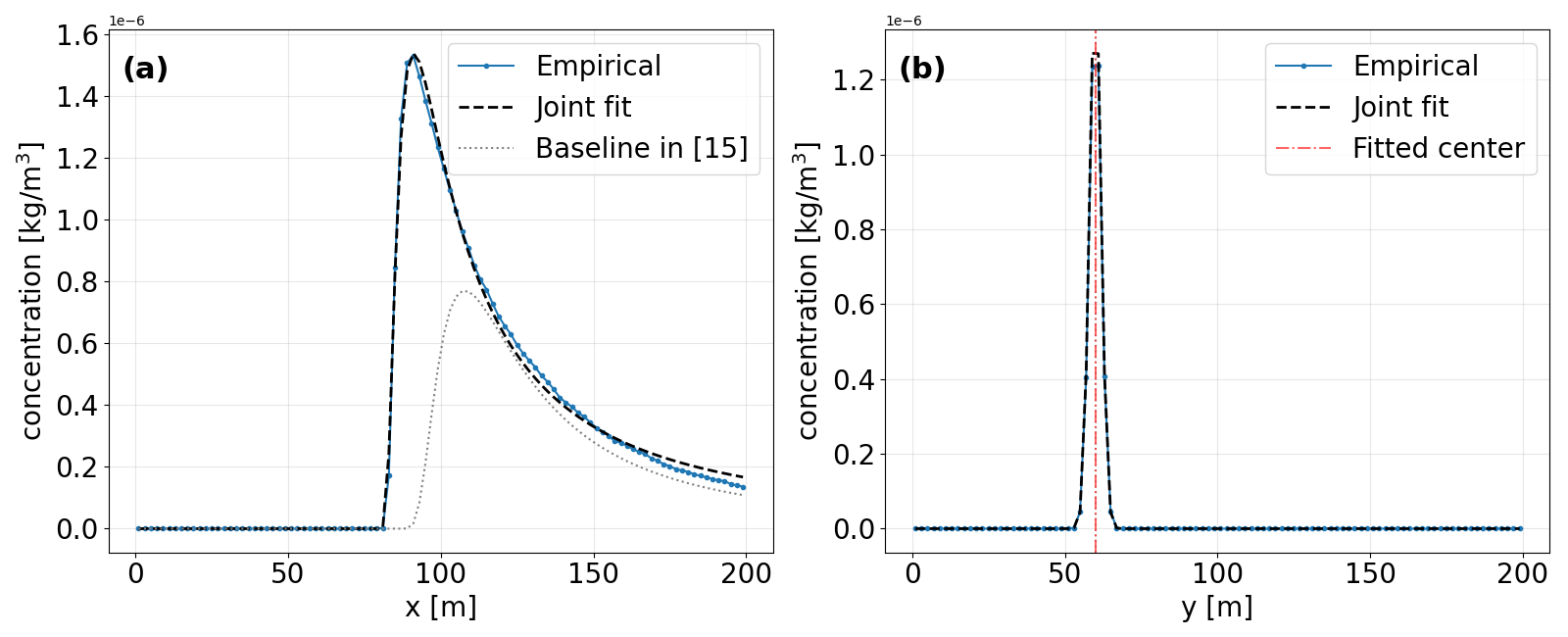}
\caption{Forward-model validation against the time-averaged MOABS field: (a) downwind profile (empirical vs.\ calibrated vs.\ uncalibrated); (b) crosswind profile at $x=100$\,m showing the fitted center $y_c$.}
\vspace{-5pt}
        \label{GP_downwind_crosswind}
    \end{minipage}
\end{figure}

\begin{figure*}[!t]
    \centering
    \includegraphics[width=\textwidth]{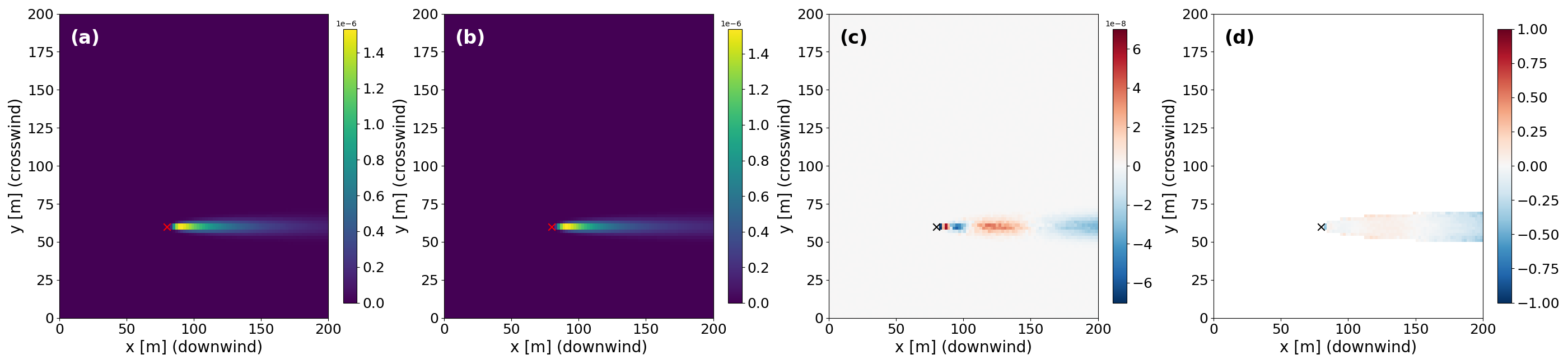}
    \caption{Spatial residual of the calibrated model vs.\ MOABS: (a) empirical field, (b) fitted model, (c) residual, (d) relative residual (masked below $1\%$ of max).}
\vspace{-5pt}
    \label{GP_residual}
\end{figure*}

\begin{figure*}[!t]
    \centering
    \includegraphics[width=0.96\textwidth]{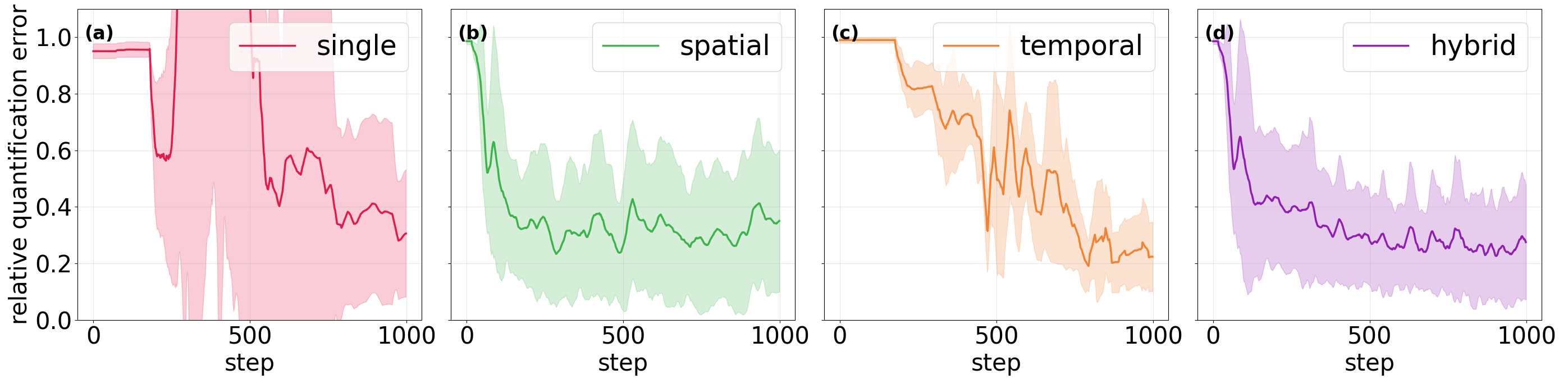}
    \caption{Per-strategy quantification error (mean $\pm$ 1 std across seeds): (a) single-agent baseline, (b) spatial, (c) temporal, and (d) hybrid planning.}
    \vspace{-5pt}
    \label{quant_err_seperately}
\end{figure*}

\subsection{Results and Discussions}
\label{sec:results_and_discussions}

\subsubsection{Forward Model Calibration}
We first verify that the joint optimization of Section~\ref{sec_time_avg_gp} yields a forward model consistent with the global time-averaged MOABS field. Fig.~\ref{GP_downwind_crosswind}(a) compares the empirical downwind profile against the calibrated model and an uncalibrated baseline using literature dispersion coefficients;
the calibrated model tracks the empirical peak and decay, while the baseline underestimates the near-source peak and mislocates the downwind maximum. Fig.~\ref{GP_downwind_crosswind}(b) examines the crosswind profile at $x=100$\,m, where the fitted center $y_c$ coincides with the empirical peak to within $0.01$\,m, confirming that the crosswind offset terms in~\eqref{eq_y_center}
correctly capture the crosswind drift in this environment. The overall fit achieves $R^2=0.998$. Fig.~\ref{GP_residual} shows the resulting spatial residual: the bulk of the domain shows negligible disagreement, with the largest relative residuals confined to a narrow region immediately downwind of the source. We do not pursue a more flexible
dispersion parameterization in this work, as our focus is on path-planning strategy rather than forward-model design.

\subsubsection{Spatial, Temporal, and Hybrid Strategies}
Fig.~\ref{quant_err_seperately} (solid line: mean, shaded band:
$\pm 1$ std across seeds, obtained by restarting the MOABS dataset
from different time offsets) and Table~\ref{tab:hybrid_numbers}
compare all four strategies. The spatial strategy's quantification
error drops earliest -- a direct consequence of the $N$ agents
covering distinct regions in parallel to localize the source
quickly, reaching the $40\%$ threshold at step 132 -- but its
steady-state band remains wide ($33.1\%\pm25.4\%$): once localized,
spatial exploration alone does not add enough information to
stabilize the estimate. The temporal strategy shows the opposite
pattern: since all $N$ agents retrace a single leader trajectory,
the fleet cannot converge until the leader locates the source,
explaining its much slower threshold crossing (step 466), but the
time-offset revisits noticeably narrow the error band and push the
mean error to a lower steady state ($25.0\%\pm13.7\%$). The hybrid
strategy, which executes the spatial warmup of Stage 1 until the
anchor converges and then switches to the temporal
sequential-following of Stage 2, inherits both strengths: it
reaches the $40\%$ threshold at step 222 -- roughly $1.7\times$
slower than spatial but more than twice as fast as temporal --
while its steady-state error ($25.3\%\pm18.8\%$) is nearly identical
to the temporal strategy's mean, with variance intermediate between
the two single-mode strategies. This confirms that switching from
spatial to temporal diversity at the point of anchor convergence
effectively balances the localization-versus-quantification
trade-off.

\begin{table}[!t]
\centering
\caption{Quantification error metrics for spatial, temporal, and
hybrid planning strategies, computed from the mean error curve
across seeds (anchor-restricted argmax).}
\label{tab:hybrid_numbers}
\begin{tabular}{lccc}
\hline
Strategy & Steps to $40\% $ error & Tail mean & Tail std \\
\hline
Spatial  & 132 & $33.1\%$ & $25.4\%$ \\
Temporal & 466 & $25.0\%$ & $13.7\%$ \\
Hybrid   & 222 & $25.3\%$ & $18.8\%$ \\
\hline
\end{tabular}
\end{table}

\subsubsection{Effect of Time-Averaging Window Length} \label{sec:window_length_result} 
\begin{figure}[!h]
    \centering
    \begin{minipage}[b]{3.5in}
        \centering
        \includegraphics[width=\linewidth]{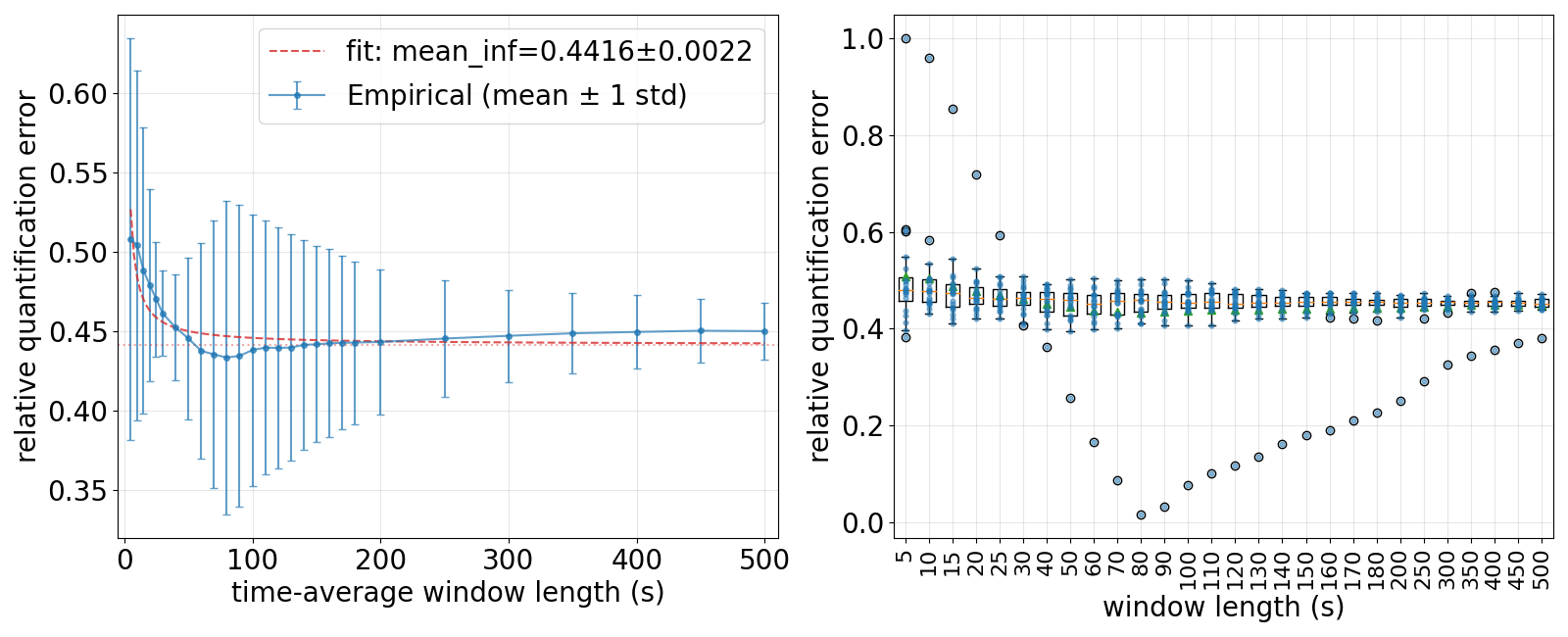}
    \caption{Relative error of $\hat{Q}$ vs.\ window length $\tau$ (5--500\,s): variance drops sharply by $\tau\approx100$\,s.}
    \vspace{-8pt}
        \label{time_avg_upper_bound}
    \end{minipage}
\end{figure}

Finally, we examine how the length of the time-averaging window
$\tau$ in \eqref{eq_time_avg} affects quantification accuracy. To
isolate this effect from path planning, $\hat Q$ is obtained
directly from the window-averaged concentration field via the
closed-form solution $\hat{Q}=\sum_i y_i f_i / \sum_i f_i^2$, where $f_i$ is the calibrated forward model's unit-rate output at grid point $i$. Sweeping $\tau$ from $5\,s$ to $500\,s$, we draw 20 non-overlapping windows per length across the MOABS dataset and
record the relative error of $\hat Q$ (Fig.~\ref{time_avg_upper_bound}).
Error variance decreases sharply from $5\,s$ to $100\,s$, as longer dwell times suppress turbulence-driven fluctuations, while the mean error shows strongly diminishing returns beyond this point (asymptotic fit: $\text{mean}_\infty = 0.4416 \pm 0.0022$, stable within $\tau\in[100,200]$). This makes indefinitely extending $\tau$ impractical; Stage 2's sequential-following mechanism instead achieves an equivalent effect without a single agent dwelling for a long, low-marginal-value period.

% \begin{figure}[!h]
%     \centering
%     \begin{minipage}[b]{3.5in}
%         \centering
%         \includegraphics[width=\linewidth]{Figures/quant_err_in_one.png}
%         \caption{Quantification error (mean ± 1 std across seeds) comparing spatial, temporal, and hybrid (uncalibrated) planning strategies.}
%         \label{quant_err_in_one}
%     \end{minipage}
% \end{figure}

% 要讲time-averaged操作和GP model fitting，
% 讲MOABS simulator
% 以及数据处理？

% MOABS filament-based仿真器介绍
% 发现并修复的两个bug：(不一定需要讲)
% Crosswind风场恒定偏置（rand()引入的+0.05m/s偏置，20亿采样点验证）及修复
% Filament释放速率质量守恒bug（ff.n vs obj(i).N）及修复，Nmax定容方法
% 修复前后对比：quant误差从~80%降至~44%，残差对称性改善（可以放修复前后的残差热力图对比）
% 说明这一节的意义：为后续策略比较提供了更可信的仿真基准

%%%%%%%%%%%%%%%%%%%%%%%%%%%%%%%%%%%%%%%

% 实验设置（域大小、源位置、seed数量等）
% spatial vs temporal vs hybrid 多seed对比：均值±方差带图（就是你们最后在调的那张图）
% 结果分析：hybrid策略是否在收敛速度和最终误差上同时优于两种单一策略；方差是否更小（更稳定）
% 可选：对残留误差来源的简要讨论（提一下forward model幂律形式的局限，但说明不是本文范围）

%%%%%%%%%%%%%%%%%%%%%%%%%%%%%%%%%%%%%%%%%%%%%
\section{Conclusion}
\label{sec:Ch5-Conclusion}
This paper presented a spatial-temporal hybrid path planning strategy for multi-UAV methane emission rate quantification, built on a time-averaged measurement model and a forward Gaussian plume model calibrated against a filament-based plume simulator. Simulation results showed that spatial diversity localizes the source and starts reducing quantification error earliest, while temporal sequential-following gives the lowest variance and the lowest steady-state relative error. The hybrid strategy switches from the former to the latter once the anchor point converges. It reached $40\%$ relative error about $1.7\times$ slower than the spatial strategy but more than twice as fast as the temporal strategy, with intermediate variance (see Table~\ref{tab:hybrid_numbers}). The window-length study also showed that lengthening the averaging window has diminishing returns, so balancing spatial and temporal diversity is more efficient than dwelling longer at each location. Future work includes validating the strategy on real field-test data and using a more flexible forward model to reduce the model mismatch.

\bibliographystyle{IEEEtran}
\bibliography{references}

\end{document}